\documentclass[a4paper,11pt]{article}
\usepackage[T1]{fontenc}
\usepackage{natbib}
\usepackage[normalem]{ulem}
\usepackage{verbatim}
\usepackage{graphicx}
\usepackage{amssymb}
\date{}
\begin{document}
\title{Stretching and squeezing of sessile dielectric drops by the optical radiation pressure}
\maketitle 
\begin{center}
\author{\Large Hamza Chra\"ibi$^{1,2}$, Didier Lasseux$^2$, Eric Arquis$^2$}\\
\author{\Large R\'egis Wunenburger$^1$ and Jean-Pierre Delville$^1$}\\
~~\\
$^1${Universit\'e Bordeaux I\\Centre de Physique Mol\'eculaire Optique et Hertzienne (CPMOH), UMR CNRS 5798\\
351 Cours de la Lib\'eration\\ 33405 Talence cedex, France.}\\
~~\\
$^2${Universit\'e Bordeaux I\\Transferts, \'Ecoulements, Fluides, \'Energ\'etique (TREFLE), UMR CNRS 8508\\
Esplanade des Arts et M\'etiers\\
33405 Talence Cedex, France.}\\
\end{center}

\begin{abstract}

We study numerically the deformation of sessile dielectric drops immersed in a second fluid when submitted to the optical radiation pressure of a continuous Gaussian laser wave. Both drop stretching and drop squeezing are investigated at steady state where capillary effects balance the optical radiation pressure. A boundary integral method is implemented to solve the axisymmetric Stokes flow in the two fluids. In the stretching case, we find that the drop shape goes from prolate to near-conical for increasing optical radiation pressure whatever the drop to beam radius ratio and the refractive index contrast between the two fluids. The semi-angle of the cone at equilibrium decreases with the drop to beam radius ratio and is weakly influenced by the index contrast. Above a threshold value of the radiation pressure, these ``optical cones'' become unstable and a disruption is observed. Conversely, when optically squeezed, the drop shifts from an oblate to a concave shape leading to the formation of a stable ``optical torus''. These findings extend the electrohydrodynamics approach of drop deformation to the much less investigated "optical domain" and reveal the openings offered by laser waves to actively manipulate droplets at the micrometer scale.\\
\end{abstract}


\section{Introduction}
\label{intro}
Since the seminal works of Zeleny \citep{zeleny17} and Taylor \citep{taylor64} on electrified jets and drops, of Pellat \citep{pellat1895} on dielectric liquid rise between vertically oriented parallel electrodes, or of Lippmann \citep{lippmann1875} on electrowetting, the control of fluid interfaces deformation under an applied field has received an increasing attention for the past century. Such a control technique plays a significant role in emerging micro/nano technologies with applications as different as electrospraying \citep{ganan-calvo04}, ink-jet printing \citep{baddie97}, electrospinning of polymer fibres \citep{hohman01a,hohman01b}, surface relief patterning \citep{schaffer00}, tuneable lensing \citep{berge00}, or microfluidics \citep{squires05}. When acting on finite volume objects such as drops, the electric stress tends to elongate the fluid interface in the direction of the electric field, as described by the well-established electrohydrostatic theory \citep{garton64,taylor64,miskis81,basaran89a,basaran89b}. This approach is usually relevant to the extreme situations where either both fluids, inside and outside the drop, are insulating dielectrics with no free charges present at their interface or the drop is a conducting fluid while the surrounding one is insulating. For a drop with zero net electric charge placed in a uniform external electric field, the electrohydrostatic theory predicts that the drop surface is always deformed into a prolate spheroid. Drop deformation toward an oblate spheroid was nevertheless observed in experiments \citep{allen62}. To explain qualitatively this different types of shape, Taylor proposed a theory based on the leaky dielectric model \citep{taylor66}, which was the basis for further developments leading to the so-called electrohydrodynamics \citep{melcher69,saville97}. A recent investigation of electric deformations on fluid interfaces \citep{shkadov02} showed that prolate and oblate shapes in the field direction can be obtained for both drops and bubbles depending on conductivity and dielectric constant ratios. A magnetic field can as well interact with fluid interfaces and form well-organized peak structures \citep{rosensweig97,lange00}, or elongate magnetic droplets \citep{bacri82,banerjee99}.

\begin{figure}[h!!!]
\begin{center}
  \includegraphics[scale=0.6]{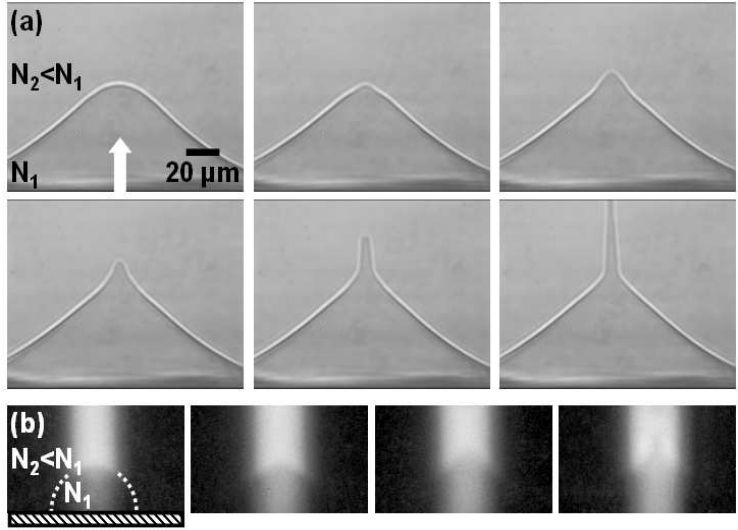}
\caption{(a) Dynamics of the draining and the deformation of a thin wetting film by the optical radiation pressure of a continuous Ar$^+$ laser beam. The initially flat interface (not shown) takes a rounded shape which further becomes near-conical and eventually destabilizes to give birth to a jet. The fluids are phase-separated liquids close to a liquid-liquid critical point (see \citep{wunenburger06a} for a detailed description of the system). The thin film results from the wetting transition occurring close to the critical point. Time delays from the first image are respectively 12~s, 18~s, 22~s, 23~s and 24~s (from top left to down right). Control parameters are $P=590$~mW, $\omega_0=4.2~\mu$m and $T-T_C=5$~K, where $T_C$ is the critical temperature. The arrow indicates the direction of propagation of the exciting beam, not seen by using a colored filter. (b) Dynamics of a heterogeneous drop growing adiabatically after an optical quench in composition inside the coexistence curve of a binary liquid mixture driven by a continuous Ar$^+$ laser (see \citep{lalaude00} for experimental details). The drop is heterogeneously nucleated on the bottom cell window and grows inside the beam. During the late-stage growth, its shape shifts gently from hemispherical to near-conical until destabilization of the interface by the radiation pressure. Time delays from the first image are respectively 202s, 266s and 312s (from left to right). Control parameters are $P=880$~mW, $\omega_0=12.4~\mu$m and $T =293.6$~K. The observed laser beam propagates upwards.} \label{fig photo cones}
\end{center}
\end{figure}

The quantitative description of these deformations also opened new horizons in fluid metrology by giving the opportunity to characterize, in a noncontact way, the mechanical properties of fluid interfaces, with a particular attention to the interfacial tension using an electric \citep{wohlhuter92,ramos94a} or magnetic \citep{flament96} field. A new application domain of these electric and magnetic manifestations is digital microfluidics, i.e. the manipulation of two-phase flows and droplets in microchannels \citep{squires05} as for instance, electric actuation \citep{link06} to manipulate flowing droplets.
However, in some particular cases, the use of electrodes or coils may become difficult, particularly for in situ characterization in open environments like oceans \citep{cinbis92} or for single microobjects such as cells \citep{wottawah05}. To overcome this difficulty, a second interface deformation approach due to radiation pressure effects, has been anticipated. For instance, it is now well established that acoustic radiation pressure can deform a fluid interface \citep{cinbis93} and even induce droplet ejection for drop-on-demand devices \citep{elrod89,meacham05}. Recently, with the emergence of nanobiotechnologies, this strategy has been successfully extended to micrometer and sub-micrometer scales using optical waves. For example, the optical deformation of soft materials in general, and of biological micro-objects in particular, brought new insights on the viscoelastic properties and the elasticity of red blood cells \citep{guck00,lee01,guck01}. The optical radiation pressure of a laser beam focused on a fluid interface is also known as an efficient contactless tool for fast metrological measurements of interfacial tension \citep{mitani02} and viscosity \citep{yoshitake05} at a microscopic scale, which is particularly appealing in the difficult cases of weak interfacial tensions and large viscosities. Thus, performing a detailed investigation of the deformation of drops by the optical radiation pressure in both the linear and nonlinear regimes of deformation, would combine the possibility of extending the fundamental and applied developments of finite volume electrohydrodynamics to the optical domain with the advantage of bringing quantitative insights on nonlinear deformation of finite size domains by laser waves, a domain which is still poorly known. This is the purpose of the present work.\\
The optical deformation of liquid interfaces was first observed by Ashkin \& Dziedzic \citep{ashkin73} using a focused laser beam to deform the water/air free surface. They experimentally demonstrated that the meniscus always bends toward the medium of lower optical index whatever the direction of propagation of the incident wave. Later on, Zhang \& Chang \citep{zhang88} observed the deformation of a water droplet illuminated by a linearly polarized pulsed laser considering two different energies (100 $mJ$ and 200 $mJ$). In these experiments, the optical absorption was assumed to be small enough to discard thermal effects. At the lower energy, Zhang \& Chang  observed an oscillation of the drop surface due to the propagation of capillary waves, a phenomenon that was further confirmed by Lai et al. \citep{lai89} and by Brevik \& Kluge \citep{brevik99}. In both studies, authors solved the problem numerically using a linear wave theory which assumes a linearization of both the flow and optical radiation pressure. When the water droplet was illuminated with the highest available energy (200 $mJ$), its front face adopted a sort of conical shape leading eventually to its disruption and the generation of micro-droplets at the tip. In order to avoid such a large amount of laser energy and access to stationary shapes instead of transient ones due to a pulsed excitation, new investigations were performed more recently by Casner \emph{et al.} \citep{casner01a,casner03a,casner03b} and Wunenburger \emph{et al.} \citep{wunenburger06a,wunenburger06b} using the interface between two liquid phases in coexistence close to their critical point. The interfacial tension between such phases being extremely small ($\sim ~10^{-7}$~N.m$^{-1}$) compared to the water superficial tension ($72$~mN.m$^{-1}$), deformations of many orders of magnitude larger than in previous experiment \citep{ashkin73,zhang88} could be easily observed using a continuous laser wave. They experimentally confirmed that an interface always bends toward the fluid of lower refractive index whatever the direction of propagation of the beam. At large beam powers, typically of the order of $1W$, interface deformations become nonlinear. Stable nipple-like shapes were observed when the laser wave is incident from the fluid of lowest optical index while propagation in the opposite direction leads to a needle-like shape, the disruption of the interface and a jetting instability driven by the total reflection of light within the deformation. As illustrated in Fig. \ref{fig photo cones}, interface disruption is still observed in such near-critical two-phase samples during (i) the deformation and the subsequent draining of a thin film (of largest refractive index) which wets the cell edges near the critical point and (ii) the adiabatic growth of a heterogeneous liquid drop (of largest refractive index too) during a liquid-liquid phase transition driven by a localized composition quench induced by a laser \citep{lalaude00}. While very different in nature, these last manifestations show analogies, at least at the level of experimental pictures, with the deformation and disruption of charged drops observed by Zeleny \citep{zeleny17} and Taylor \citep{taylor64} under strong electrical fields. Indeed, by increasing the electric field, drops deformation evolves from rounded to conical shapes emitting eventually a jet of micro-droplets at the tip. A theoretical study was proposed by Taylor \citep{taylor64} indicating that the stable static solution for a conductive drop submitted to both electric and capillary couplings is near-conical with a semi-angle of $49.3^{\circ}$. Since then, many experimental \citep{oddershede00,reznik04,fernandez04,chen05} and theoretical \citep{wohlhuter92,ramos94a,saville97,fernandez92} works have been performed, to analyse stability and disruption of these conical shapes as well as on the value of the semi-angle versus electric properties of the liquids (see for instance Fig.2 in Stone et al. \citep{stone98} for the dielectric case and the review of Fernandez de la Mora \citep{fernandez92} for conducting droplets). Note finally, that deformed interfaces with conical shape can be found in the absence of electromagnetic excitation. A first example is the near-conical shapes observed during the deformation of the interface between immiscible fluids by selective withdrawal, with a straw whose tip is suspended above the unperturbed interface \citep{case07,cohen02}. Other fluid dynamics examples are drop breakup from a nozzle \citep{eggers97} or sink flows in the presence of an interface \citep{courrech06}, thus illustrating the emergence of a sort of robust and general topological transition \citep{zhang04} of fluid interface deformations under localized forcing.
Motivated by some amazing effects of the radiation pressure illustrated in Fig. \ref{fig photo cones} and suspecting that finite size effects may promote the observed near-conical shapes, we propose a numerical investigation of optical deformation of sessile liquid drops by continuous laser waves. Indeed, beyond the simple deformation of spherical interfaces, we demonstrate in the present work that optical stretching can lead to near-conical shapes. Above a radiation pressure threshold, drop disruption is also observed numerically, in the absence of total reflection of light however, demonstrating a new phenomenon of interface instability. The optical analogue of oblate deformation by an electric field, induced here by an optical squeezing, is investigated as well. At large radiation pressures, the local squeezing leads to the formation of torus-like shapes. Section II is devoted to the physical model used to predict the deformation of drops by the optical radiation pressure. Section III briefly summarizes the numerical algorithm used here, based on the Boundary Integral Element Method (BIEM). Results on drop deformations for both stretching and squeezing are presented and discussed in Section IV.

\section{Physical model}

\begin{figure}[h!!!]
\begin{center}
  \includegraphics[scale=0.5]{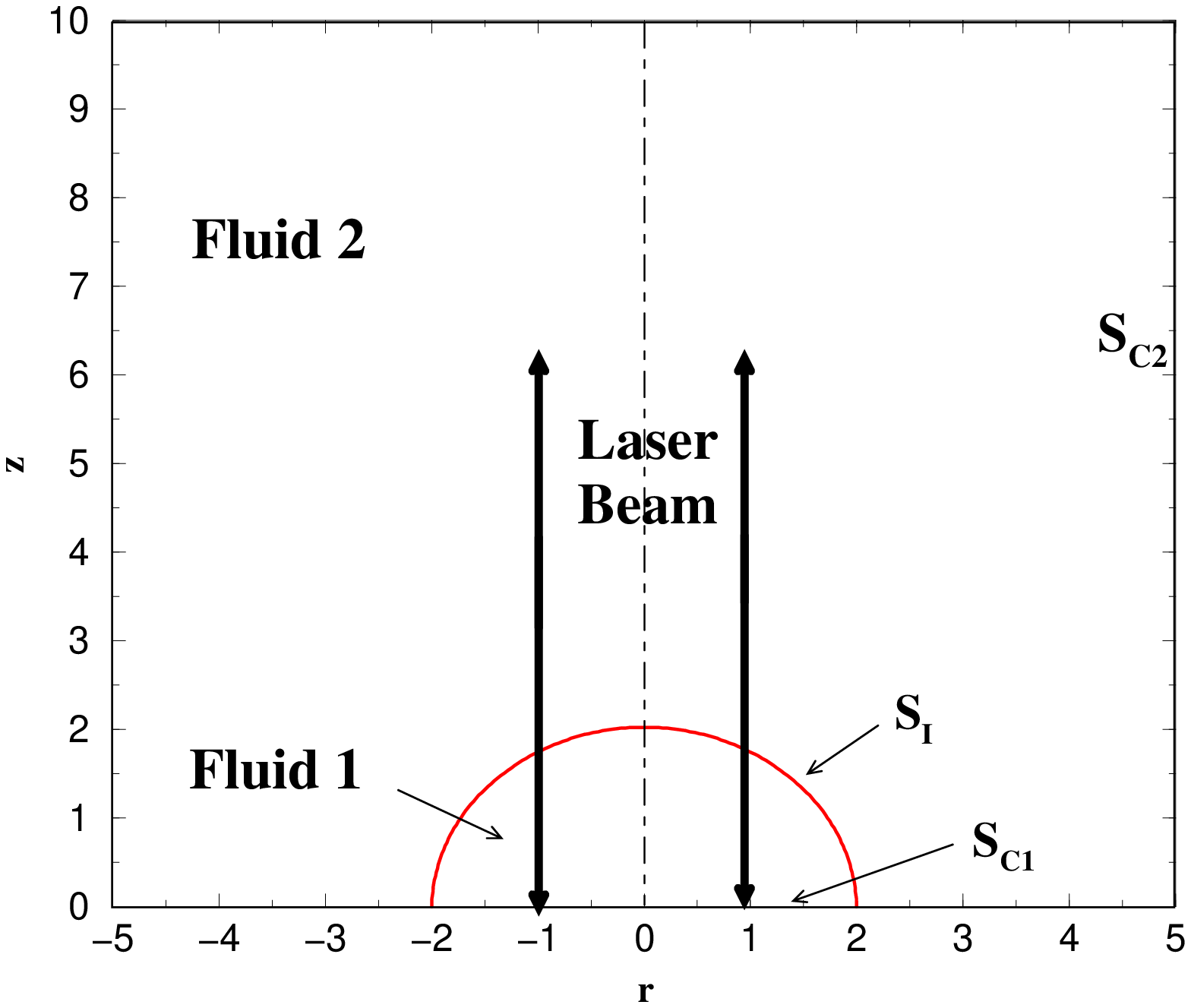}
\caption{(Color online)Schematic representation of a captive drop (fluid 1) immersed in a second liquid (fluid 2) and submitted to the optical radiation pressure of a laser beam centered on the drop axis and initially focused at the interface. $S_{C1}$, $S_{C2}$, and $S_I$ respectively denote the solid boundary with fluid 1, fluid 2 and the interface between the two fluids. The initial drop is hemispheric. The three-phase contact line is assumeed to be pinned. Lengths are made dimensionless using the beam waist $\omega_0$.} \label{schema}
\end{center}
\end{figure}

The droplet configuration under consideration, together with the notations used throughout this work, are represented in Fig. \ref{schema}.\\

To describe the hydrodynamics of the laser/fluid interaction, let us first consider a Gaussian continuous laser wave, of beam waist $\omega_0$, and make dimensionless all lengths involved in the problem using $\omega_0$. This laser wave is supposed to impinge on the interface between a semi-spherical dielectric viscous drop (marked as fluid 1) of dimensionless radius $a$ ($a=\frac{R_d}{\omega_0}$ where $R_d$ is the drop radius, $a=2$ in Fig. \ref{schema}) and a surrounding dielectric viscous fluid (marked as fluid 2) of dimensionless horizontal and vertical extensions $R$ and $H$. We assume that the three-phase contact line is motionless. Since the intensity distribution of the laser beam, centered on the drop, is supposed to be axisymmetric as in usual situations, cylindrical coordinates ($e_r$, $e_z$, $e_\phi$) with their origin O located at the center of the drop are used throughout this work so that any point \textbf{x} is located by $(r,z,\phi)$ in this reference frame of coordinates. Governing equations are written in a dimensionless form using (i) the laser waist $\omega_0$ as the characteristic length scale, (ii) the viscous relaxation velocity $u^*=\frac{\gamma}{\langle \mu \rangle}$ as the reference velocity associated to the characteristic timescale $t^*=\frac{<\mu>\omega_0}{\gamma}$ and, (iii) the reference pressure $p_i^*= \frac{\mu_iu^*}{\omega_0}~,~i=1,2$ to rescale the pressure $p_i$ in each phase. Here, $\gamma$ is the interfacial tension between fluids 1 and 2, $\mu_i$  is the dynamic viscosity of fluid i and $\langle \mu \rangle=\frac{\mu_1+\mu_2}{2}$ is the average viscosity of the fluids. 
It could be pointed out that, depending on the fluids and/or the large beam powers under consideration, (those given in Fig. 1, for instance), temperature effects may disturb or even overcome the mechanical effect of light on fluid interfaces, thus making a coupled heat and momentum transfer description necessary. The first expected additional effect is a direct laser heating due to the optical absorption of the drop and/or of the surrounding fluid. Fluids must indeed be transparent at the used optical wavelength as it is generally the case for classical liquids in the visible window. Typically, the optical absorption of water, used in Ashkin \& Dziedzic and Zhang \& Chang experiments \cite{ashkin73,zhang88}, is of the order of $3.10^{-3}cm^{-1}$ in the visible region, while that of the micellar phases used in the examples illustrated in Fig. 1 is $3.10^{-4}cm^{-1}$. In the latter, the overheating induced by a beam power of the order of 1 W is smaller than 0.1 K \citep{chraibi07}. Direct laser heating effects can then be discarded, even in critical fluids as far as temperature is not too close to the critical one. Thus, we consider in the following all liquid properties ($\gamma$, $\mu_i$, $\rho_i$, $N_i$) as constant in the presence of laser light, $\rho_i$ and $N_i$ being respectively the density and optical refraction index of fluid i. A second coupling is the thermocapillary effect. Since the interfacial tension $\gamma$ is a function of the temperature, local laser heating may drive interfacial tension gradients inducing stresses on the drop interface and its subsequent deformation \citep{loulergue81}. A typical value of $\displaystyle{|\gamma^{-1}(\frac{\partial\gamma}{\partial T})|}$ for classical fluids is $10^{-3}K^{-1}$ \cite{sammarco99}, which leads to negligible thermocapillary effects considering the above mentioned laser overheating. For the case illustrated in Fig. 1, we already found that thermocapillary interface deformation is negligible near the critical point \cite{chraibi07}. Consequently, we can safely discard temperature effects without affecting the generality of our purpose. Finally, when getting very close to the critical point, the capillary length $\displaystyle{l_C=\sqrt{\frac{\gamma}{\Delta\rho g}}}$, where $\Delta\rho$ is the density contrast and $g$ the acceleration due to gravity, vanishes while thermal fluctuations increase. Then, the interface roughness $\displaystyle{l_T=\sqrt{\frac{k_B T}{\gamma}}}$ \cite{aarts04}, where $k_B T$ is the thermal energy, increases too and may dominate capillary effects. Using the critical data given in Reference \cite{chraibi07} for deformations presented in Fig. 1, $l_C>>l_T$ requires $T-T_C>>3.10^{-2}K$, a condition which has always been fulfilled in experiments. Conversely, for classical fluids, $\gamma$ lies between $10$ and $100 mN/m$ and $\Delta\rho$ is about $10^2$ to $10^3 kg/m^3$. As a consequence, $l_T$ is orders of magnitude smaller than $l_C$ so that thermal fluctuations are also negligible for this class of fluids.

In addition, we assume in this study that inertial and gravity effects are negligible at the micrometric scale, which implies that the Reynolds and Bond number are small compared to unity. Along with the condition $l_C>>l_T$ which is also automatically satisfied as indicated above, we consider an incompressible quasi-static Stokes flow in each phase.

Therefore, the hydrodynamics of each liquid phase is described by the Stokes and mass conservation equations, respectively given by:
\begin{equation}
{\bf 0}=-\nabla q_{i}+\nabla ^{2}{\bf u_{i}}~,~i=1,2,  \label{moment}
\end{equation}
and
\begin{equation}
\nabla \cdot {\bf u_{i}}=0~,~i=1,2,  \label{mass}
\end{equation}
where ${\bf u}_i$ and $q_i$ are respectively the dimensionless velocity and the pseudo-pressure in fluid $i$. This pseudo-pressure contains the electrostrictive contribution of light in dielectrics \citep{landau60} which was demonstrated to have no incidence on the shape or the height of the deformed interface \citep{lai89,brevik99,chraibi07}.
This pseudo-pressure is defined as
\begin{equation}
q_{i}=\frac{1}{p_{i}^{\ast }}\left( p_{i}-\frac{\epsilon _{0}}{2%
}E_{i}^{2}\rho _{i}\frac{\partial \epsilon _{i}}{\partial \rho _{i}}\right),
\end{equation}
where $\epsilon_i=N_i^2$  represents the relative dielectric permittivity of fluid i, $\epsilon_0$  is the permittivity of vacuum and $E^2_i$ is the quadratic magnitude of the electric field in fluid $i$ averaged over an optical period.
Considering the classical expression of the divergence free hydrodynamic stress tensor ${\bf T}_i$,
\begin{equation}
{\bf T_{i}}=-q_{i}{\bf I}+(\nabla {\bf u_{i}}+^{t}\nabla {\bf %
u_{i}}),
\end{equation}
we can write the boundary condition at the interface as follows:
\begin{equation}
\frac{2}{1+\lambda}(\lambda {\bf T_{1}}\cdot {\bf n}-{\bf T_{2}}\cdot {\bf n})\cdot
{\bf n}=\kappa (r)-\Pi (r),  \label{stressjump}
\end{equation}
where ${\bf n}$ is the unit vector normal to the interface directed from fluid 1 to fluid 2, and $\lambda=\frac{\mu_1}{\mu_2}$ is the viscosity ratio. Eq. (\ref{stressjump}) simply expresses the fact that the normal stress on the interface is balanced by capillary forces and the optical radiation pressure, respectively represented in dimensionless forms by $\kappa(r)$ and $\Pi(r)$. Indeed, in the right hand side of Eq. (\ref{stressjump}), $\kappa(r)$ represents the dimensionless double mean curvature of the interface, given by:
\begin{equation}
\kappa (r)=\frac{1}{r}\frac{d}{dr}\left( \frac{rz^{\prime }}{\sqrt{%
1+z^{\prime 2}}}\right),
\end{equation}
where $z'=\frac{dz}{dr}$ is the local slope of the interface. The second term, $\Pi(r)$, is the contribution of the dimensionless optical radiation pressure at the drop interface. As the interface deformation is experimentally found to be directed toward the fluid of smallest refractive index whatever the direction of propagation \citep{ashkin73,casner01a}, we consider the Minkowski point of view which states that the photon momentum in a dielectric medium varies linearly with the refractive index \citep{campbell05}; for a review on the Abraham-Minkowski controversy see for instance the review of Brevik \citep{brevik79}. The amplitude of $\Pi(r)$ is nevertheless affected by the direction of propagation through its dependence on the incidence and transmitted angles. In addition, $\Pi(r)$ also depends on the polarization of the laser wave. In the present work, we preserve the axial symmetry of the laser/fluid interaction by assuming a circular polarization of the laser wave. Finally, for the sake of simplicity, we assume refringence of light at the drop interface at any incidence angle. By doing this, we eliminate situations where total reflection of light may occur at the interface as those presented in Fig. \ref{fig photo cones} for example. Therefore, the laser wave is supposed to propagate from the optically less dense fluid. Note that this choice is not restrictive at all; by removing this assumption, we would find the same type of drop deformations but with an asymmetry in amplitude due to the nonlinear behavior of the transmission and reflection Fresnel coefficients with the angle of incidence \citep{casner03a,chraibi07}. Another reason for this choice is made clear in Sec. IV.2 where a new type of laser-induced interface instability is advanced, the one demonstrated previously being triggered by total reflection of light.

On the one hand, to stretch a drop by light, and thus deform the interface outward (Sec. IV.1), the wave propagates downward. In this case, one has $N_1>N_2$, and the dimensionless expression of the optical radiation pressure is $\Pi(r)=\Pi^{dn}(r)\frac{\omega_0}{\gamma}$ where $\Pi^{dn}(r)$ is given by:
\begin{equation}
\Pi ^{dn}(r)=\frac{I(r)}{c}\cos \theta _{i}(2N_{2}\cos \theta
_{i}-\psi^{dn}(N_{2}\cos \theta _{i}+N_{1}\cos \theta _{t})).
\label{raddown}
\end{equation}%
On the other hand, to squeeze a drop and thus deform the interface inward (Sec. IV.3) we choose $N_1<N_2$ and consider a beam propagating upward. In this case the expression of the rescaled optical radiation pressure becomes  $\Pi(r)=\Pi^{up}(r)\frac{\omega_0}{\gamma}$ where $\Pi^{up}(r)$ is given by:
\begin{equation}
\Pi ^{up}(r)=-\frac{I(r)}{c}\cos \theta _{i}(2N_{1}\cos \theta
_{i}-\psi^{up}(N_{1}\cos \theta _{i}+N_{2}\cos \theta _{t})).
\label{radup}
\end{equation}
In Eqs (\ref{raddown}-\ref{radup}), we have respectively denoted by $\theta_i$ and $\theta_t$ the incidence and transmission angles; $c$ is the light celerity in vacuum. One has $\theta_i=\arctan(z')$ while $\theta_t$ depends on the direction of propagation. One has $\theta_t=\arcsin(\eta \sin\theta_i)$ or $\theta_t=\arcsin(\frac{1}{\eta} sin\theta_i)$ respectively for upward and downward propagations, where $\eta=N_1/N_2$ is the refractive index ratio. Moreover, $I(r)$ represents the intensity of the Gaussian laser beam. By neglecting the weak z-dependence of the beam radius, its expression is given by:
\begin{equation}
I(r,z)\approx I(r)=\frac{2P}{\pi \omega _{0}^{2}}e^{-2r^{2}},
\end{equation}
where $P$ is the beam power. In addition, $\Psi^{up}$ and $\Psi^{down}$ are the Fresnel
transmission coefficients of energy fluxes. These coefficients are obtained
from the ratio of the transmitted to incident normal components of
Poynting's vectors and are expressed as
\begin{equation}
\Psi^{up}=\Psi^{down}=\frac{2N_{1}N_{2}\cos \theta _{i}\cos \theta
_{t}}{(N_{1}\cos \theta _{i}+N_{2}\cos \theta _{t})^{2}}+\frac{%
2N_{1}N_{2}\cos \theta _{i}\cos \theta _{t}}{(N_{2}\cos \theta
_{i}+N_{1}\cos \theta _{t})^{2}}
\end{equation}
for a circulary polarized beam. In order to quantify effects of the laser wave on the interface deformation,
it is convenient to define the electromagnetic to Laplace pressure ratio.\ This ratio  $%
\xi $, taken at $r=0$ ($\theta _{i}=\theta _{t}=0)$, is defined as:
\begin{equation}
\xi =\left\vert \Pi {(r=0)}\right\vert =\frac{4P}{\pi c\omega _{0}\gamma }%
\frac{N_{i} \left\vert  N_{2}-N_{1}\right\vert }{(N_{2}+N_{1})}~,~i=1,2,
\end{equation}
where $N_i$ refers to the optical index of the incidence fluid. When no slip is assumed at the interface, along with the fact that fluids are immiscible, it follows that the velocity ${\bf u}$, of the interface $S_I$ is equal to that of each fluid particle on $S_I$, i.e.:
\begin{equation}
{\bf u(x)}={\bf u_{1}(x)}={\bf u_{2}(x)}{ \ \ \ \ \ }{\bf x}
\in S_{I}
\end{equation}
Moreover, the movement of the interface is described using a Lagrangian approach. It consists in following each fluid particle of the interface in its Lagrangian motion according to the kinematic condition:
\begin{equation}
\frac{d{\bf x}}{dt}={\bf u}({\bf x}){ ~~~ }{\bf x}\in S_{I}.
\label{kin}
\end{equation}%
This condition indicates that the interface is advected along with the flow until the
equilibrium is reached for which normal velocities along the interface are
zero, i.e. ${\bf u(x).n}=0$, ${\bf x}\in S_{I}$.
Finally, we assume a classical no-slip boundary condition on all the solid boundaries of the domain
\begin{equation}
{\bf u_{i}(x)=0}{ \ \ \ \ \ \ \ }{\bf x}\in S_{Ci},{ }i=1,2.
\end{equation}%
The above system of equations is solved using a Boundary Integral Element Method (BIEM). Due to the axial symmetry of the laser/drop interaction, it consists in an axisymmetric integral formulation making use of the fundamental solution of Stokes equations. The solution is sought with a constant boundary elements discretization technique according to the numerical scheme described below.
\section{Numerical algorithm}
A brief description of the numerical algorithm is presented in this section.
For more extensive details on the BIEM applied to a two-phase axisymmetric
flow, the reader may refer to the review by Tanzosh et al. on the solution
of free surface flow problems using this technique \citep{tanzosh92}. The
BIEM reveals to be an excellent tool to solve interfacial flow problems with
high resolution as reported in the analysis of flow involving electric and
magnetic fields \citep{sherwood87} or buoyancy \citep{manga94} \citep{koch94}%
.\newline
Because solutions to Stokes equation can be formulated in terms of Green's
functions, we can rewrite the governing equations as a system of integral
equations over the boundaries of the computational domain. When doing so,
the boundary integral form of the Stokes equation for fluid $i$ ($i=1,2$)
can be written as follows \citep{pozrikidis92}\newline
\begin{equation}
\frac{1}{2}{\bf u}_{i}{\bf (x)}=\int_{S_{I}+S_{Ci}}{\bf U}\cdot ({\bf T}_{i}%
{\bf \cdot n}_{i})dS_{y}-\int_{S_{I}}{\bf n}_{i}{\bf \cdot K\cdot u}%
_{i}dS_{y},
\end{equation}%
${\bf n}_{i}$ being the unit normal vector directed toward the outside of
fluid domain $i$. In this last expression, ${\bf U}$ and ${\bf K}$ are
second and third order tensors forming the Green's kernel for velocity and stress associated to the Stokes equation.\ These two tensors are
respectively given by \citep{pozrikidis92}:
\begin{eqnarray}
{\bf U(d)} &=&\frac{1}{8\pi }(\frac{1}{d}{\bf I}+\frac{{\bf dd}}{%
d^{3}}), \\
{\bf K(d)} &=&-\frac{3}{4\pi }(\frac{{\bf ddd}}{d^{5}}).
\end{eqnarray}%
In these two relationships, ${\bf d=x-y}$, ${\bf y}(r_{y},z_{y})$
is the integration point. Once boundary conditions on $S_{I}$, $S_{C1}$ and $S_{C2}$ are used, the two-phase Stokes problem can be written in the following compact form:
\begin{eqnarray}
\lefteqn{{\bf u(x)}=\int_{S_{I}}{\bf U}\cdot
{\bf n}(\kappa (r_{y})-\Pi (r_{y}))dS_{y}+}  \nonumber \\
&&\frac{2}{1+\lambda } [ (1-\lambda )\int_{S_{I}}{\bf n.K.}{\bf u}dS_{y}+\lambda \int_{S_{C1}} {\bf U\cdot (T_{1}\cdot n)}dS_{y}-  \nonumber \\
&& \int_{S_{C2}}{\bf U\cdot (T_{2}\cdot n)}dS_{y}].  \label{zebigone}
\end{eqnarray}
Here, $S_{I}$, $S_{C1}$ and $S_{C2}$ must be understood as axisymmetric
surfaces instead of their trace in the plane of Fig. \ref{schema}. In Eq. (\ref{zebigone}), the first term on the right hand side describes the flow
contribution from surface tension and radiation pressure, whereas
the second term accounts for the shear rate contrast at the interface. As
expected, this second term vanishes when there is no viscosity contrast between the
two phases ($\lambda =1$). The third and fourth terms account for the shear
occurring on $S_{C1}$ and $S_{C2}$ as a result of the no-slip boundary
condition.\newline
The solution of the problem consists in the computation of the velocity, $%
{\bf u}$, on the interface as well as the stress over all the boundaries $%
S_{I}$,$S_{C1}$ and $S_{C2}$.\ This is performed once Eq. (\ref{zebigone}) has been discretized using boundary elements.\ Here, we use
constant boundary elements for which pressure, velocity, and hence stress,
take constant values on each element, equal to that at the central node.\
The overall numerical procedure can be summarized as follows. (1) All boundaries $S_{I}$, $S_{C1}$ and $S_{C2}$ are discretized
with line segments of total number $N$. The fluid-fluid interface $S_{I}$ is
parameterized in terms of an arc length, $s$, so that the double mean
curvature can be accurately computed using the following expression:
\begin{equation}
{\kappa (r)=}\left\vert \frac{d{\bf t}}{ds}\right\vert {+\frac{z^{\prime
}(r)}{r(1+z(r)^{\prime }{}^{2})^{1/2}}},
\end{equation}%
where ${\bf t}$ is the tangential vector to $S_{I}$. The number of mesh points on $S_{I}$ is $70$ for a typical computation. Each horizontal and vertical solid
boundaries are meshed using approximately $50$ uniformly distributed points.
An increase in the mesh resolution for the interface and solid boundaries did
not show any significant change in the results. (2) At each time step, the azimuthal integration of the integrals arising once Eq. (18) is discretized with N constant boundary elements is performed analytically \citep{lee82}, \citep{graziani89} reducing Eq. (\ref{zebigone}) to line
integrations which are finally performed using Gauss quadratures %
\citep{davis84}. Elliptic integrals resulting from the azimuthal integration
are evaluated using power series expansions \citep{bakr85}. (3) Once all integrals are calculated, the linear system  ${\bf Aw=b}$
is formed where ${\bf w}$ is the vector of the unknown boundary values of
velocities on $S_{I}$ and normal stress and shear on $S_{I}$, $S_{C1}$ and $%
S_{C2}.$\ The matrix ${\bf A}$ results from boundary integrals computed
with the procedure detailed above.\ The vector ${\bf b}$ is obtained from
the product of boundary integrals and the known part of the normal stress
jump on $S_{I}$ as it appears in the first term in the right hand side of
Eq. (\ref{zebigone}). This linear system is solved at each time step
using a direct method based on a LU decomposition.\ This step provides $%
{\bf u(x)}$ on $S_{I}$. (4) Finally, the motion of the interface is captured using the
kinematic condition (Eq. \ref{kin}) and an explicit first-order Euler time
scheme the discrete form of which can be written as follows:
\begin{eqnarray}
r(t+\Delta t) &=&r(t)+u_{r}(t)\Delta t,  \\
z(t+\Delta t) &=&z(t)+u_{z}(t)\Delta t,
\end{eqnarray}%
where $\Delta t$ is the time step. This time step is typically chosen to be
about $20$ times smaller than $t ^{\ast}$. Once the interface has been
moved (stretched or squeezed), the mesh is reconstructed using a smoothing procedure
with local cubic splines.
Starting from an initial semicircular interface at rest, the laser beam is switched
on at $t=0$ and the interface starts to deform towards the fluid of smallest
refractive index. The overall algorithm is repeated until final equilibrium
is reached and this is achieved when
\begin{equation}
{\bf u(x).n}<\epsilon { \ \ \ \ \ \ }{\bf x}\in S_{I},
\end{equation}%
where $\epsilon <<1$ is a user fixed parameter.
\section{Results}
In this section, we first present results on drop stretching by the optical radiation pressure. An outward deformation of the liquid drop is obtained when its optical index of refraction is larger than that of the surrounding fluid (i.e. $N_1>N_2$), resulting in the formation of near-conical shapes at different dimensionless initial drop radii $a$ and refractive index ratios, characteristic of most dielectric fluids. An instability arising from the destabilization of the drop above a threshold value of $\xi$ is also demonstrated. Results on drop squeezing corresponding to $N_1<N_2$ are presented in a second part. In this case, drop deformation becomes concave and may reach a torus-like shape.
\subsection{Drop stretching by the optical radiation pressure}
\begin{figure}[h!!!]
\begin{center}
  \includegraphics[scale=0.65]{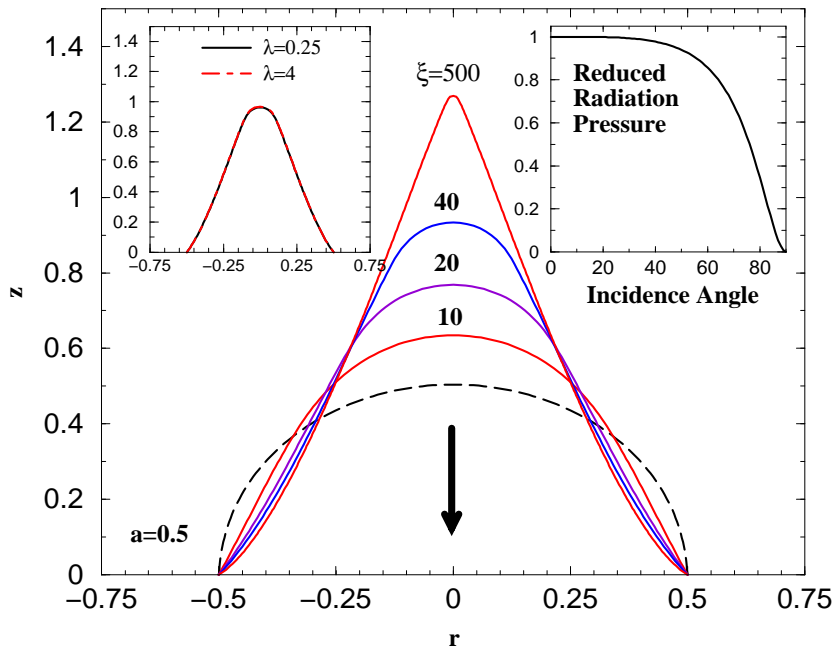}
\caption{(Color online)Optical stretching of a drop of dimensionless radius $a=\frac{R_d}{\omega_0}=0.5$ ($R_d$ is the drop radius, $\omega_0$ the beam waist). The number associated to each outward deformation refers to the pressure ratio $\xi$. The initial spherical shape is represented by the dashed line. The arrow indicates the direction of propagation of the laser beam. Left Inset: steady shapes at $\xi=50$ for two viscosity ratios showing no effect of $\lambda$ on the steady-state solution. Right Inset: variation of the optical radiation pressure rescaled by its value at normal incidence, versus the incidence angle for $\frac{N_1}{N_2}=1.1$. } \label{cone.5}
\end{center}
\end{figure}

Simulations of outward deformations were performed with $\lambda=\frac{\mu_1}{\mu_2}=2$ and $\eta=\frac{N_1}{N_2}=1.1$. This value of the index ratio, corresponding for example to a toluene droplet immersed in water, was chosen as a representative couple for the water/oil interface. The generality of the investigation is nonetheless totally preserved because $\lambda$ has no influence on the stationary shape of deformed drop (see the left Inset of Fig. \ref{cone.5}) and no bifurcation has been observed in our study. Variations in $\eta$ just lead to qualitative differences in optical radiation pressure effects, as illustrated in the near-conical shape study (see Sec. IV.2). Stationary shapes of an optically stretched liquid drop of dimensionless radius $a=0.5$ are illustrated in Fig. \ref{cone.5} for different optical to Laplace pressure ratios $\xi$.
In Fig. \ref{cone.5}, one clearly see that the shape of the drop progressively varies from rounded ($\xi$=10, 20, 40) to near-conical ($\xi$=500). These stationary shapes result from the competition between capillary forces and optical radiation effects. Increasing $\xi$ increases the height of the deformation and thus induces in turn an increase of the curvature of the interface. This effect is maximized at the tip of the deformation due to the Gaussian profile of the light intensity and the normal incidence at the interface at $r=0$, as illustrated in the right inset of Fig. \ref{cone.5} . Therefore, an important increase in $\xi$ eventually leads to pointed drop shapes, assuming the fixed contact line hypothesis and the finiteness of the drop volume. Although the stress on the interface is here of optical origin rather than electrically or magnetically induced, the deformation of drops by the optical radiation pressure shows striking similarities with those observed in electro- and magneto-hydrodynamics \citep{wohlhuter92,ramos94a,stone98,sherwood87}. There are nevertheless two major differences. First, the relative extension of the exciting field versus the drop size, considered as infinite in the case of electric or magnetic fields, is intrinsically finite due to the Gaussian shape of the laser intensity in the current analysis. Moreover, the associated radiation pressure nonlinearly decreases with the incident angle through the Fresnel transmission coefficient, drastically reducing the mechanical effect of the beam at the edge of the deformation when its aspect ratio increases significantly (see the right Inset of Fig. \ref{cone.5}). Second, the pointed shape observed on the axis does not originate from a local increase of the electric field at the tip as in the electrostatic case because at a given $\xi$, the optical radiation pressure at the tip, where the incidence angle is close to zero, remains regular.
\begin{figure}[h!!!]
\begin{center}
  \includegraphics[scale=0.55]{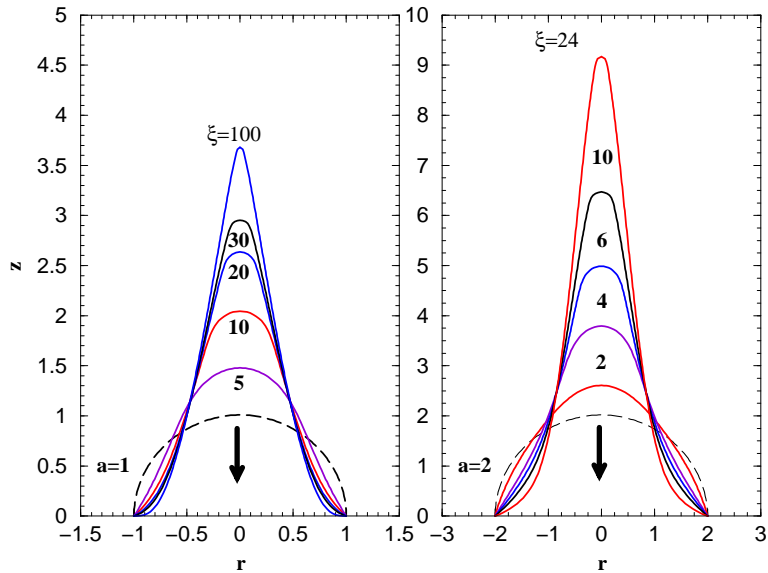}
\caption{(Color online)Optical stretching of a drop of dimensionless radius $a=1$ (left) and $a=2$ (right). The number associated to each outward deformation refers to the pressure ratio $\xi$. The initial spherical shape is represented by the dashed line. The arrow indicates the direction of propagation of the laser beam.} \label{cone1-2}
\end{center}
\end{figure}

\begin{figure}[h!!!]
\begin{center}
  \includegraphics[scale=0.65]{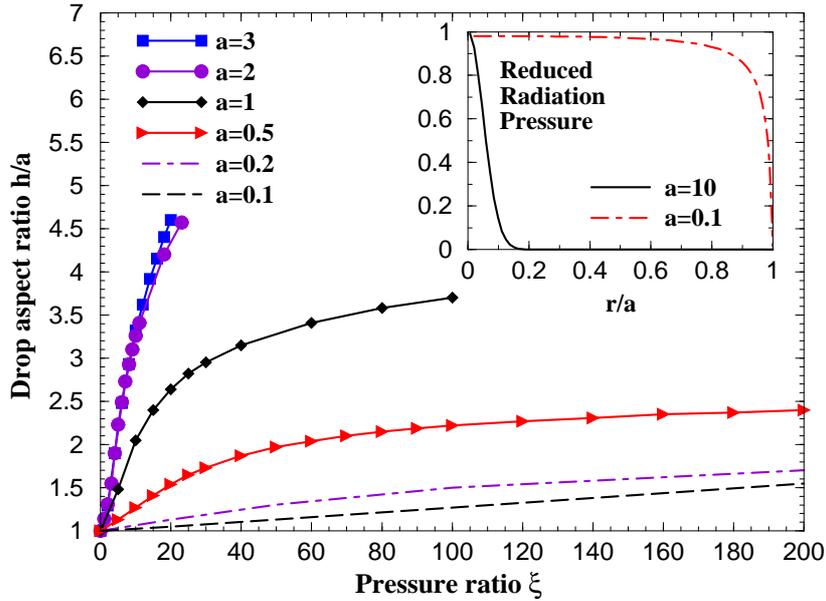}
\caption{(Color online)Variation of the reduced drop stretching h/a versus the pressure ratio $\xi$ for different dimensionless drop radii $a$. The inset shows the variation of the optical radiation pressure rescaled by its value at normal incidence versus $r/a$.} \label{h_xi_ext}
\end{center}
\end{figure}

Fig. \ref{cone1-2} shows the outward deformation of a drop with dimensionless radius $a=1$ and $a=2$. Qualitatively, droplet deformations are very similar to those of Fig. \ref{cone.5}. Increasing the dimensionless radius of the drop leads to a decrease of the $\xi$ value required to achieve a given dimensionless deformation height. In other words, a larger drops will have a higher aspect ratio $h/a$ than a smaller one for the same $\xi$, $h$ being the equilibrium height defined as $h=z(r=0,t\rightarrow\infty)$. The $\xi$ variation of the drop aspect ratio $h/a$, is illustrated in Fig. \ref{h_xi_ext} for different drop radii $a$. The fact that the $h/a(\xi)$ curves depend on $a$ can be explained by the distribution of radiation pressure on the drop, as shown in the inset of Fig. \ref{h_xi_ext}. When $a\ll1$, the electromagnetic intensity applied to the drop is almost uniform. The radiation pressure mainly depends on the incidence angle which vary for an initial spherical drop from $0^{\circ}$ at the tip to $90^{\circ}$ at the contact line. So, it is nearly uniform except near the contact line where it decreases down to zero, as shown in the inset of Fig. \ref{h_xi_ext}. The associated electromagnetic normal stress is almost uniform all along the drop surface. Conversely, when $a \gg 1$, the radiation pressure mainly depends on the gaussian intensity of the laser beam, as the incidence angles near the tip remain close to zero. The electromagnetic normal stress is localized on the top of the drop. Consequently, the drop is more deformed for $a \gg 1$ than for $a \ll 1$.

\subsection{Optical cone formation and interface instability}
\begin{figure}[h!!!]
\begin{center}
  \includegraphics[scale=0.55]{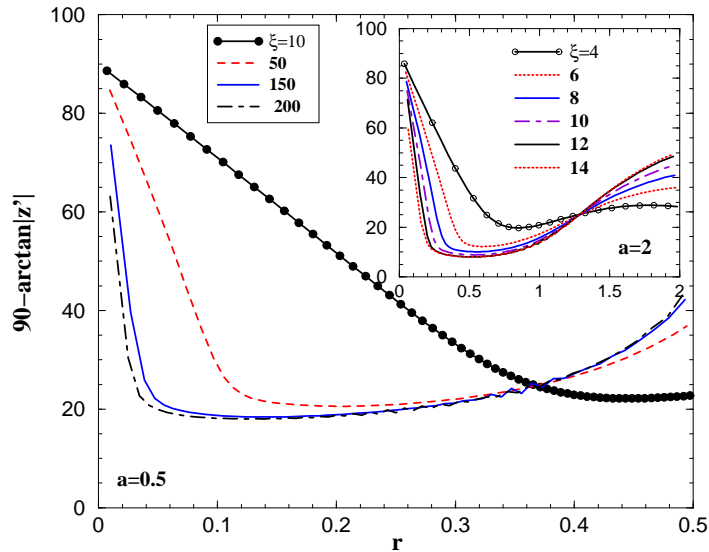}
\caption{(Color online)Variation of the angle between the drop axis and the tangent at the interface of a deformed drop versus $\xi$ for $N_1/N_2=1.1$ and for both $a=0.5$ and $a=2$ (Inset).} \label{paliers}
\end{center}
\end{figure}

\begin{figure}[h!!!]
\begin{center}
  \includegraphics[scale=0.55]{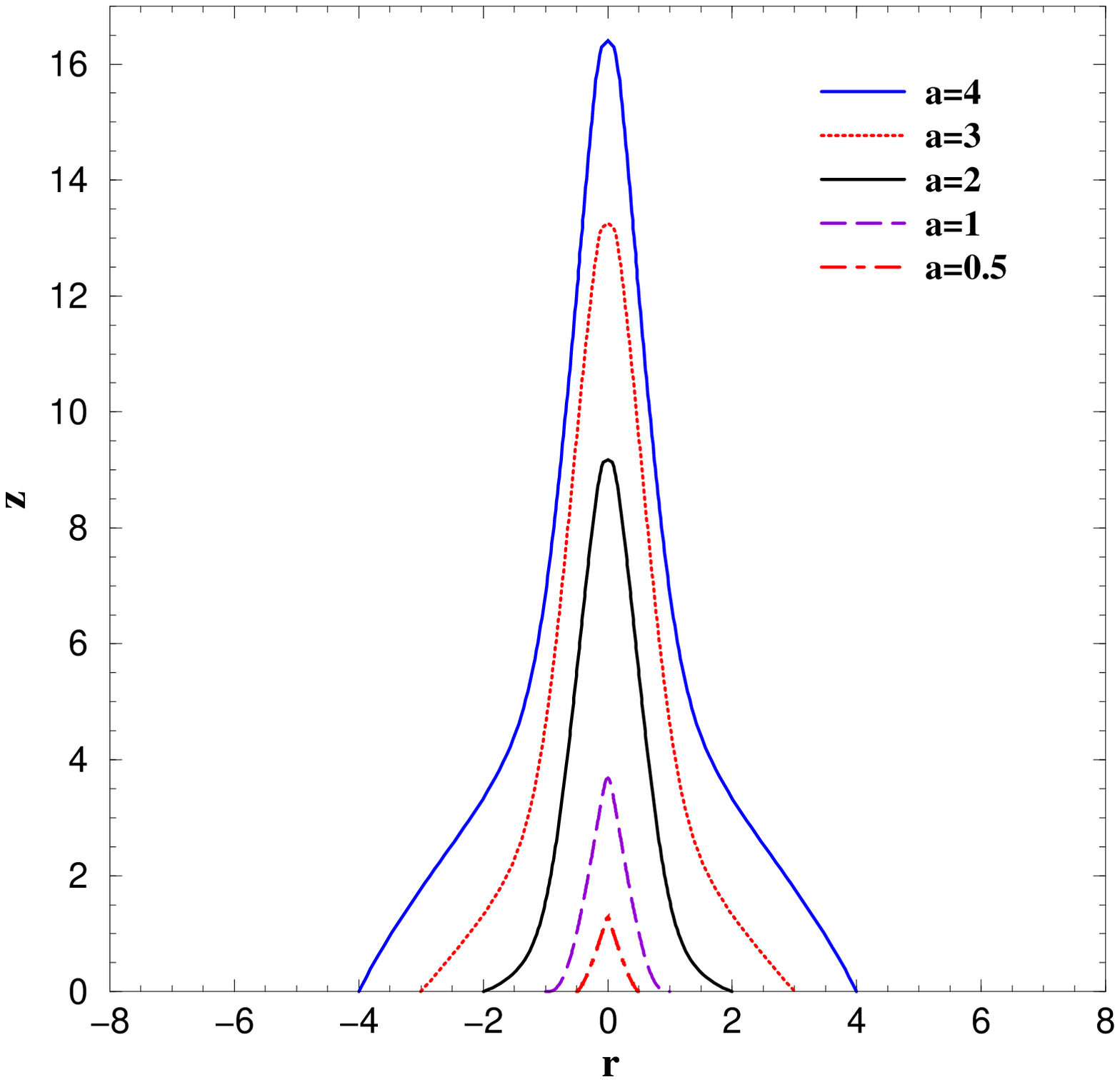}
\caption{(Color online)Stationary tip shapes at $N_1/N_2=1.1$ for different dimensionless radii at $\xi$ values above the semi-angle threshold emergence $\xi_P$.} \label{cones}
\end{center}
\end{figure}

\begin{figure}[h!!!]
\begin{center}
  \includegraphics[scale=0.55]{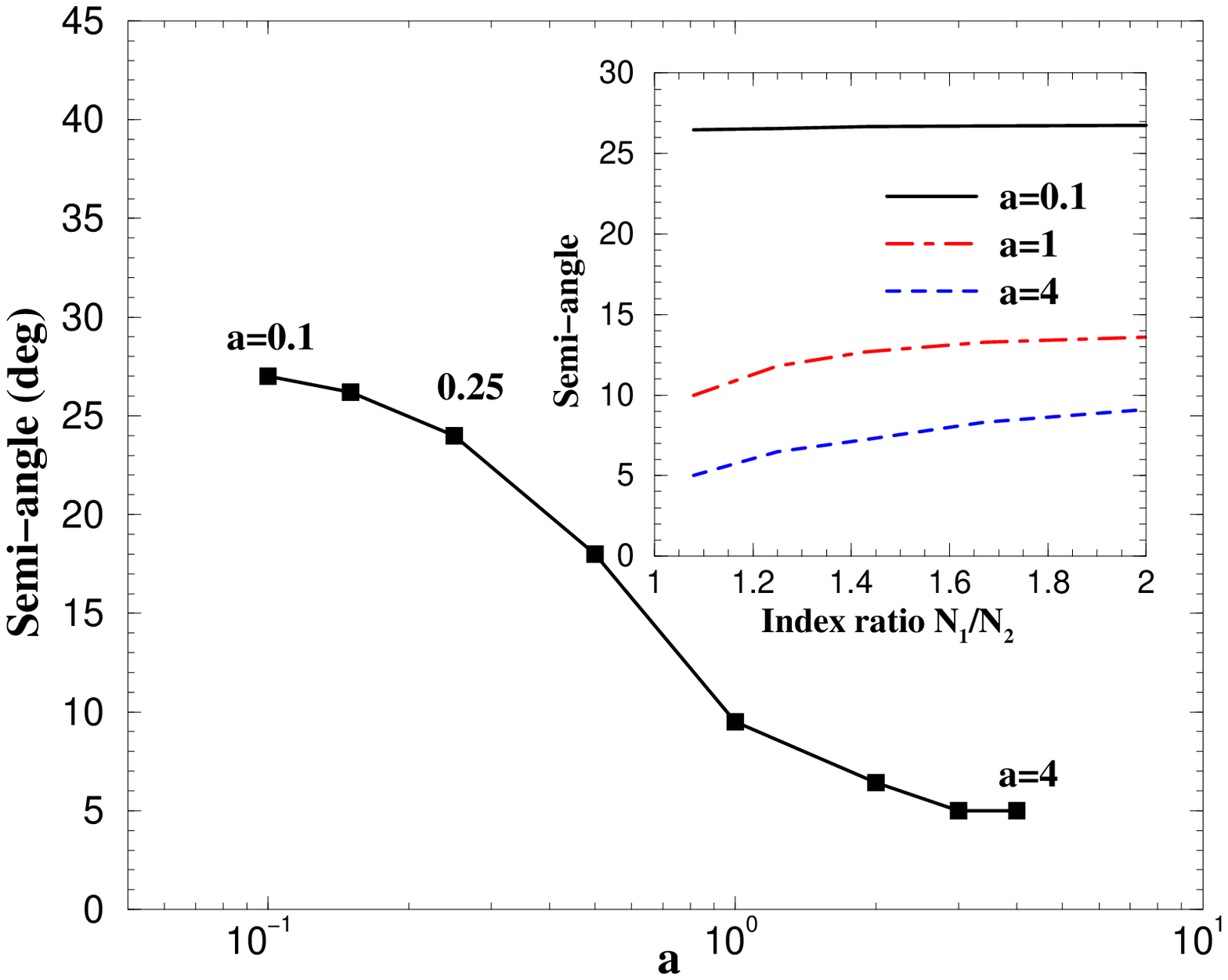}
\caption{(Color online)Semi-logarithmic variation of the semi-angle of near-conical stationary shapes versus the dimensionless drop radius $a$ ($\eta=1.1$). Inset: variations of the semi-angle versus the refractive index ratio over the range of investigated dimensionless drop radii.} \label{angles}
\end{center}
\end{figure}

As illustrated in Figs. \ref{cone.5} and \ref{cone1-2}, drop deformations with near-conical shape seem to emerge when increasing the pressure ratio $\xi$ whatever the beam radius. It could be objected that such a generic shape requires an optical radiation pressure behaving as $1/r$ to balance the Laplace pressure, while the optical coupling shows a dependence in both $\exp(-r^2)$ and a nonlinear function of the incident angle. However, as experimentally illustrated in Fig. \ref{fig photo cones} and in the following, under certain conditions, conical shapes can approximate the true deformation profile with a high degree of accuracy. The existence of such type of deformation is still surprising if we take into account predictions for dielectric drops subjected to electric fields \citep{wohlhuter92,ramos94a,stone98,sherwood87} which indicate that conical shapes cannot exist for dielectric constant ratio smaller than 20. Here, the situation is nevertheless different since the beam profile is inhomogeneous and the deformation efficiency depends on the beam extension. To quantitatively investigate the emergence and the existence of cone-like shapes, we implemented the following procedure. For a cone, the semi-angle is defined as $90-\arctan(|z'|)$ (in $^{\circ}$), where $z'$ is the slope of the generating line. Considering a drop of dimensionless radius $a$, we plot $90-\arctan(|z'|)$ for increasing $\xi$. A cone-tip signature appears through the emergence of a plateau, corresponding to a constant local slope $z'$ along the interface, which defines its semi-angle. As illustrated in Fig. \ref{paliers} for $a=0.5$ and $a=2$, i.e. for beam waist respectively larger and smaller than the droplet radius, a plateau emerges at a threshold $\xi_P \simeq 50$ ($a=0.5$) and $\xi_P \simeq 8$ ($a=2$) for $N_1/N_2=1.1$. Its wideness increases for $\xi>\xi_P$ while the corresponding semi-angle remains almost constant. This saturation of the semi-angle at large $\xi$, observed at any investigated dimensionless drop radius, demonstrates the robustness of the near-conical deformation. Results in Fig. \ref{paliers} also suggest that the value of the semi-angle varies with the dimensionless drop size a.
In Fig. \ref{cones}, we present stationary deformations of liquid drops of different dimensionless radii when $\xi > \xi_P$.
A near-conical shape is systematically observed. These results also illustrate, in a different representation, the variation of the vertical semi-angle of the cone versus the radius of the drop. To be more quantitative, Fig. \ref{angles} shows that the semi-angle (i) remains finite at large values of $\xi$ whatever the value of $a$ and (ii) exhibits two different asymptotic values at large and at small $a$, the larger one, at small $a$, logically corresponding to the steep variation of the mean slope of the deformation $h/a(\xi)$ shown in Fig. \ref{h_xi_ext}, and the smaller one at large $a$ to the smooth variation of $h/a(\xi)$. For small $a$, the asymptotic value is roughly $26^{\circ}$ while it is close to $5^{\circ}$ at large values of $a$.

The finiteness of the asymptotic value of the cone semi-angle at large $\xi$ is the most striking behavior of this class of laser induced interface deformations. The fact that the optical radiation pressure pulls the cone tip with a diverging strength when $\xi \longrightarrow \infty$ would make one indeed think that the cone slope also diverges. This behavior has a consequence on the cone stability. In fact, above a second threshold value $\xi_I$ of $\xi$, the drop deformation becomes unstable leading to an interface breakup, as illustrated and discussed below.

In the $a \ll 1$ limit, the experimental conditions used to deform the drop are similar to those used with electric fields in capacitors since the radial extension of the exciting laser wave is significantly larger than the drop size. Moreover, the asymptotic value of the semi-angle $26^{\circ}$ is close to the minimum one, $30^{\circ}$, found for dielectrics in electric fields. This result is surprising since the dielectric constant ratio required to reach this minimum angle is $(N_1/N_2)^2=1.21$ while in the electrostatic case it corresponds to $17.6$. In addition, the asymptotic value of the semi-angle in the case of optical excitation at small $a$ was found to remain equal to $26^{\circ}$ for $(N_1/N_2)^2=17.6$, an unrealistic optical situation for dielectric liquids. This mismatch between electrical and optical excitation is likely due to the fact that the mechanisms involved in the formation of near-conical shapes are different. However, even if we do not have yet any quantitative explanation, these observations, at least at the level of experimental observations, suggest the existence of a possible common phenomenology of the deformation of dielectric drops by electric and optical excitation. To give another insight on this appealing comparison, we analyzed the index contrast variation of the semi-angle for different dimensionless drop radii (Inset of Fig. \ref{angles}). The refractive index ratio was varied from $1.05$ to $2$, this last alue corresponding to an upper limit for the free surface of dielectric liquids (the largest index of refraction we know is that of diiodomethane $CH_2I_2$, $N=1.7425$ at $20^{\circ}C$). Given a dimensionless radius $a$, the measured angle weakly increases with $N_1/N_2$ when $a=1$, while it is almost constant for $a<<1$. This last behavior can be explained by the two following facts: (i) the laser incidence experiences all angles between $0^{\circ}$ and $90^{\circ}$ when the beam waist is larger than the drop radius and (ii) the light intensity is almost uniform over the drop when $a<<1$.

\begin{figure}[h!!!]
\begin{center}
  \includegraphics[scale=0.55]{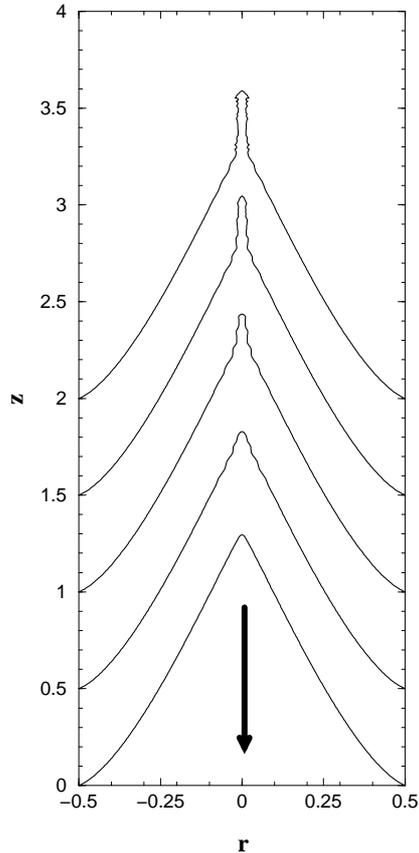}
\caption{Dynamics of the disruption of the near-conical shape of a drop of dimensionless radius $a=0.5$ at different times (the reduced time step is 0.02). The lowest profile is the stable stationary shape obtained for $\xi=500$. At $\xi=700$, the destabilization of the deformed drop occurs. The corresponding profiles (from down to top) have been translated for clarity. The arrow indicates the direction of propagation of the laser beam.} \label{jet}
\end{center}
\end{figure}

All the near-conical shapes presented above were obtained before the previously mentioned instability threshold. They were thus stable. Given a dimensionless drop radius a, when the pressure ratio $\xi$ reaches what we called the second threshold value $\xi_I$ (the first one, $\xi_P$, was attached to the emergence of a near-conical shape), a destabilization of the deformation is observed, leading to a disruption of the interface at the tip, as illustrated in Fig. \ref{angles}. A refinement in mesh and time steps did not show any change on this behavior. Even if the optical radiation pressure is known to be able to destabilize a fluid interface and form a liquid jet \citep{casner03a,wunenburger06b}, this numerical result is quite different because jetting was previously induced by total reflection of light within the deformation when the laser beam propagates from the fluid having the largest refractive index towards that of lowest refractive index, as it is the case in Fig. \ref{fig photo cones}. To prevent this effect, here the opposite situation was chosen. The beam propagates from the fluid having the lowest refractive index towards that of largest refractive index, a situation where ``nipple-like'' interface deformations are generated at large beam powers for fluids of infinite extensions \citep{casner03b} and for which further increase of the power keeps the nipple like shape stable. The main difference with the present investigation is that now, the volume of one of the two fluids, the drop, becomes significantly small compared to the other one. Increasing $\xi$ still increases the height of the deformation, but the drop turns to adopt a near-conical steady shape, the semi-angle being surprisingly preserved along a wide plateau which extension increases towards smaller values of $r$ while increasing $\xi$ (see Fig. \ref{paliers}). This complex balance between the Laplace pressure and the laser incidence dependent radiation pressure is preserved up to $\xi_I$, as illustrated in Fig. \ref{jet}. The saturation of the semi-angle at large $\xi$ observed numerically implies that the radiation pressure applying along the cone slope linearly increases with $\xi$ whereas the cone shape does not evolve anymore: this qualitatively explains why a destabilization of the strongly deformed drop occurs beyond the threshold value $\xi_I$, as illustrated in Fig. \ref{jet}. The quantitative investigation of this drop disruption under a very high optical radiation pressure regime and the subsequent disruption, which clearly deserve a devoted study, will be presented in future development.
\subsection{Drop squeezing by the optical radiation pressure}
\begin{figure}[h!!!]
\begin{center}
  \includegraphics[scale=0.55]{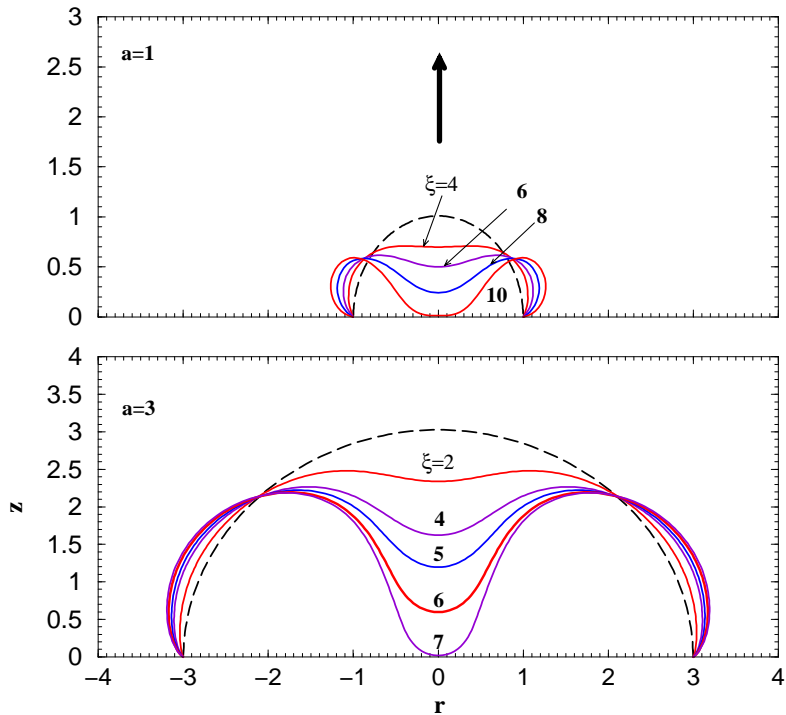}
\caption{(Color online)Optical squeezing of drops of dimensionless radius $a=1$ and $a=3$. The number associated to each inward deformation refers to the pressure ratio $\xi$. The initial spherical shape is represented with a dashed line. The arrow indicates the direction of propagation of the laser beam.} \label{int2-3}
\end{center}
\end{figure}

\begin{figure}[h!!!]
\begin{center}
  \includegraphics[scale=0.55]{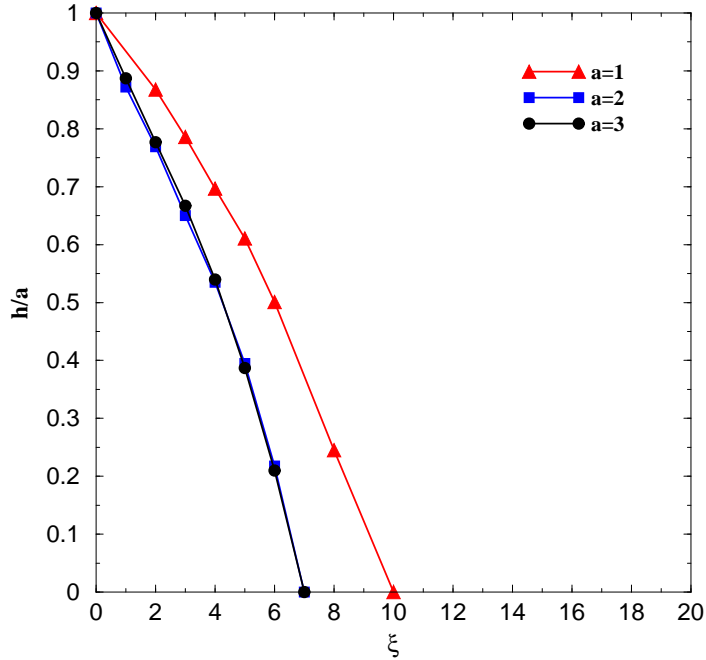}
\caption{(Color online)Variation of the reduced drop squeezing $h/a$ versus the pressure ratio $\xi$ for different dimensionless drop radii $a$. } \label{h_int}
\end{center}
\end{figure}

In this section, we investigate the deformation of a liquid drop of refractive index lower than that of the surrounding liquid. We preserved $\lambda=\frac{\mu_1}{\mu_2}=2$ for calculations and reverse the refractive index contrast by taking $\eta=\frac{N_1}{N_2}=0.9$ in order to force a drop squeezing instead of stretching. Moreover, we restricted our investigation to the case $a \geq 1$ in order to avoid any significant effect of multiple reflection inside the droplet. As before, the generality of the purpose is preserved because the value of $\lambda$ has no influence on the stationary shape of deformed drops and variations in $\eta$ just modify quantitatively the optical radiation pressure effects. Fig. \ref{int2-3} shows the resulting variations of the stationary shape of a liquid drop of dimensionless radius $a=1$ and $a=3$ at various pressure ratio $\xi$. The spherical shape flattens at low values of $\xi$. For larger beam power, the curvature reverts and the shape becomes concave. At large $\xi$, the concavity reaches the solid boundary and flattens more and more in its central region.
As in the stretching case, the main difference between the deformations obtained at different dimensionless drop radii is that the amplitude of the deformation at a given $\xi$ decreases with the drop radius and reaches some asymptotic behavior. In Fig. \ref{h_int} we have represented the evolution of the reduced height of the drop $h/a$ as a function of the pressure ratio $\xi$ for drop radii $1 \leq a \leq 3$.
For $a\geq2$, the evolution of the reduced height shows little change while increasing $a$. A behavior similar to that was already observed for drop stretching. These observations can as well be explained by the decrease of the incidence angle range of the illuminated area of the drop while increasing its radius. For $a=1$, larger radiation pressure is required to obtain the same amplitude of the interface deformation. The former argument proposed for drop stretching can be used again to explain the observed increase in $\xi$ required to reach equilibrium: when $a>>1$, the local illumination of the drop leads to larger deformations compared to the case where the field is more uniform ($a=1$).
While optical stretching leads to near-conical shapes at large $\xi$, the squeezing of a drop should give birth to the formation of a stable ``optical torus''. Indeed, by assuming a fixed contact line on the substrate, we implicitly prevent drop spreading, thus promoting the formation of an annular rim. The film thinning at the center should as well induce local ``dewetting'' when the deformation reaches the substrate, as in the electric case \citep{yeo07}. Here, we assume that molecular forces are much larger than any external forcing at a nanometric scale and, speculate that Van der Waals forces will dominate the final stage of the deformation \citep{seemann01}, a result which could be retrieved by including the disjoining pressure into the Stokes equations or boundary conditions at the interface \citep{yeo07}. Consequently, even if the formation of stable torus from radiation pressure effects is still speculative and deserves an experimental demonstration, the present investigation shows that lasers allow for drop squeezing with morphologies which were, to the best of our knowledge, never even suggested while using electric fields.

\section{Conclusions}
Although many appealing applications of optically induced drop deformations can be advanced, among them, the optical stretcher tool which was developed by Guck et al. \citep{guck01} to deform red blood cells and discriminate between sane and cancerous cells \citep{lincoln04} or the contactless viscoelastic micro-characterization of fluids \citep{sakai01}, very few theoretical or numerical studies were performed in this field, especially in the nonlinear regime of deformation which is strongly influenced by finite-volume effects. The objective of this paper was thus to provide new elements for understanding such type of drop deformation and go even further in order to illustrate the specificities of drop deformation by the optical radiation pressure. Both deformation cases, the prolate one when the drop is stretched and the oblate one when it is squeezed, were studied at steady state as a function of the amplitude of the optical radiation pressure, normalized by the Laplace pressure, and for varying drop radii.
We found that the elongation of stretched drops significantly varies with the beam waist and can adopt a near-conical shape at large optical radiation pressures, as suggested by the experimental illustrations of Fig. \ref{fig photo cones}. Contrary to the classical electrodynamics case, where a minimal dielectric constant ratio is required to reach cone shapes \citep{stone98}, these shapes are observable at any optical refractive index ratio. The semi-angle was found to be a decreasing function of the drop radius, showing two asymptotic values. When the drop is much smaller than the beam, the semi-angle is close to that obtained for electrically deformed dielectric drops. Above a threshold in radiation to Laplace pressure ratio, a disruption of the drop is observed. This behavior is appealing because even if a similar phenomenon was already observed experimentally by Zhang \& Chang \citep{zhang88} on drops, Casner \emph{et al.} \citep{casner03a} on extended two-phase fluids and presented in Fig. \ref{fig photo cones} for wetting films and growing drops, the experimental conditions were totally different in the sense that instability was triggered by the total reflection of light within the deformation. Here, the beam incidence has been chosen to precisely avoid this mechanism, showing that an optically stretched drop can still become unstable above a radiation pressure threshold due to finite volume effects. We also investigated drop squeezing versus the amplitude of the radiation to Laplace pressure ratio, and for various drop radii. At large radiation pressure, the drop shape shifts from oblate to concave. As in the stretching case, the squeezing significantly varies with the beam waist. With a further increase of the radiation pressure, the concavity reaches the solid boundary giving birth to a stable torus-like shape.
Finally, by extending the developments on finite volume electrohydrodynamics to the optical domain, our approach raised numerous questions on analogies and differences in drop deformations control by electromagnetic fields in general. Not all of these questions received definite answers. The optical jetting, for instance, could open new horizons in microdroplet dispensing. Consequently, even if the formation of stable cones and torus from radiation pressure effects deserve experimental investigation, the present numerical study illustrates the opportunities offered by laser waves to actively manipulate droplets at the micrometer scale and reach dynamic and stable intriguing nonlinear drop morphologies.

\textbf{Acknowledgements}\\
This research was supported by the Centre National de la Recherche Scientifique (France) and the Conseil R\'egional d'Aquitaine under contract 20040205003A.\\

\bibliographystyle{apsrev}

\end{document}